\documentclass[twocolumn,times,tighten]{aastex631}

\shorttitle{Spectral properties of MAXI J0637$-$430}
\shortauthors{Thomas et al.}
\hypersetup{linkcolor=blue,citecolor=blue,filecolor=cyan,urlcolor=red}
\usepackage{amsmath}
\usepackage{float}
\usepackage{mwe}
\usepackage[caption=false]{subfig}

\submitjournal{The Astrophysical Journal}
\received{: 22-10-2021}
\accepted{: 2-12-2021}
\published{: 3-2-2022}
\usepackage{scalerel}
\usepackage{tikz}
\usetikzlibrary{svg.path}

\definecolor{orcidlogocol}{HTML}{A6CE39}
\tikzset{
  orcidlogo/.pic={
    \fill[orcidlogocol] svg{M256,128c0,70.7-57.3,128-128,128C57.3,256,0,198.7,0,128C0,57.3,57.3,0,128,0C198.7,0,256,57.3,256,128z};
    \fill[white] svg{M86.3,186.2H70.9V79.1h15.4v48.4V186.2z}
                 svg{M108.9,79.1h41.6c39.6,0,57,28.3,57,53.6c0,27.5-21.5,53.6-56.8,53.6h-41.8V79.1z M124.3,172.4h24.5c34.9,0,42.9-26.5,42.9-39.7c0-21.5-13.7-39.7-43.7-39.7h-23.7V172.4z}
                 svg{M88.7,56.8c0,5.5-4.5,10.1-10.1,10.1c-5.6,0-10.1-4.6-10.1-10.1c0-5.6,4.5-10.1,10.1-10.1C84.2,46.7,88.7,51.3,88.7,56.8z};
  }
}

\newcommand\orcidicon[1]{\href{https://orcid.org/#1}{\mbox{\scalerel*{
\begin{tikzpicture}[yscale=-1,transform shape]
\pic{orcidlogo};
\end{tikzpicture}
}{|}}}}
\usepackage{hyperref}

\begin{document}
\shortauthors{Thomas et al.}
\title{Spectral properties of soft X-ray transient MAXI J0637$-$430 using AstroSat}
\correspondingauthor{Shivappa B. Gudennavar}
\email{shivappa.b.gudennavar@christuniversity.in}

\author{Neal Titus Thomas \orcidicon{0000-0001-9460-3264}}
\affiliation {Department of Physics and Electronics, CHRIST (Deemed to be University) \\
Bangalore Central Campus, Bengaluru - 560029, India}
\nocollaboration{1}
\author{Shivappa B. Gudennavar \orcidicon{0000-0002-9019-9441}}
\affiliation {Department of Physics and Electronics, CHRIST (Deemed to be University) \\ Bangalore Central Campus, Bengaluru - 560029, India}
\nocollaboration{1}
\author{Ranjeev Misra}
\affiliation {Inter-University Centre of Astronomy and Astrophysics \\ Post Bag 4, Ganeshkhind, Pune - 411007, India}
\nocollaboration{1}
\author{Bubbly S. G. \orcidicon{0000-0003-1234-0662}}
\affiliation {Department of Physics and Electronics, CHRIST (Deemed to be University) \\
Bangalore Central Campus, Bengaluru - 560029, India}
\nocollaboration{1}

\begin{abstract}
\footnotesize
\noindent
Soft X-ray transients are systems that are detected when they go into an outburst, wherein their X-ray luminosity increases several orders of magnitude. These outbursts are markers of the poorly understood change in the spectral state of these systems from low/hard state to high/soft state. We report the spectral properties of one such soft X-ray transient: MAXI J0637$-$430, with data from the \textit{SXT} and \textit{LAXPC} instruments on-board \textit{AstroSat} mission. The source was observed for a total of $\sim$ 60 ks over two observations on 8$^{th}$ and 21$^{st}$ November, 2019 soon after its discovery. Flux resolved spectral analysis of the source indicates the presence of a multi-colour blackbody component arising from the accretion disk and a thermal Comptonization component. The stable low temperature ($\sim$ 0.55 $keV$) of the blackbody component, points to a cool accretion disk with an inner disk radius of the order of a few hundred $km$. In addition, we report the presence of a relativistically broadened Gaussian line at 6.4 $keV$. The disk dominated flux and photon power law index of $\gtrapprox 2$ and a constant inner disk radius indicate the source to be in the soft state. From the study we conclude that MAXI J0637$-$430 is a strong black hole X-ray binary candidate.
\end{abstract}


\section{Introduction} 
\label{sec:introduction}
\noindent
Soft X-ray transients are a subclass of low mass X-ray binaries (LMXBs) that appear as extremely faint sources ($L = 10^{30} - 10^{33} $~$erg~s^{-1}$) during most of their lifetime. They are characterized by a non-steady transfer of mass onto the compact object and they occasionally undergo sporadic outbursts, which occur at intervals of 1 - 60 years \citep{Chen1997,Tetarenko2016}. This causes their X-ray luminosity to increase by a factor of upto $10^{7}$ \citep{vanParadijs1995} which then decays back to quiescence with an e - folding timescale of $\sim 30$ days \citep{Chen1997}. During the short outbursts, they emit enough X-rays, which makes them the brightest X-ray sources ($L = 10^{37} - 10^{38} $~$erg~s^{-1}$) in the sky. The occurrence of these outbursts is attributed to instabilities in the accretion disk that are both viscous and thermal in nature \citep{Meyer1981,Cannizzo1995,King1998,Lasota2001}. Soft X-ray transients, especially the ones harbouring a black hole, usually go undetected; and are discovered only when they undergo an outburst. 
\\[6pt]
MAXI J0637$-$430 is one such source, which was first detected by the \textit{MAXI/GSC} nova alert system during seven scan transits from $2^{nd}$ $-$ $3^{rd}$ November, 2019 in the $2 - 4$ $keV$ and $4 - 10$ $keV$ bands. The scans revealed the source to be located at RA (J2000) = 06 h 38 m 54 s, Dec (J2000) = $-$42 h 45 m 57 s in the soft band ($2 - 4$ $keV$) and at RA (J2000) = 06 h 37 m 43 s, Dec (J2000) = $-$43 h 03 m 15 s in the hard band ($4 - 10$ $keV$) with 90\% confidence level \citep{Negoro2019}. Since its first detection, MAXI J0637$-$430 underwent a considerable increase in its flux from $59$ $\pm$ $6$ mCrab to $\sim 200$ mCrab in the $2 - 4$ $keV$ band and from $32$ $\pm$ $6$ mCrab to $\sim 50$ mCrab in the $4 - 10$ $keV$ band \citep{Negoro2019}. A Target of Opportunity (ToO) observation performed on $3^{rd}$ November, 2019 by the \textit{Swift} mission detected MAXI J0637$-$430 at RA (J2000) = 06 h 36 m 23.59 s, Dec (J2000) = $-$42 h 52 m 04.1 s, which was consistent with MAXI's hard band localization. Its X-ray spectrum modelled using absorbed disk blackbody + power law with inner disk temperature ($kT_{in}$) of  0.9 $\pm$ $0.1$ $keV$ and power-law index ($\Gamma$) of $2.3$ $\pm$ $0.8$ indicated that the source underwent an outburst and transitioned from hard to soft spectral state. An optical counterpart with a brightness of u = $14.87$ $\pm$ $0.02$ (Vega) was detected by the \textit{UVOT} instrument on-board \textit{Swift} at RA (J2000) = 06 h 36 m 23.23 s Dec (J2000) = $-$42 h 52 m 04.25 s. Since there are no known stars at this position, it was deemed that this optical source underwent a significant brightening, as is common for the optical counterpart of black hole low mass X-ray binaries (BH-LMXBs) during outbursts \citep{Kennea2019}. Follow up observation by the \textit{NuStar} mission found the source to be at a flux of $\sim$ 95 mCrab. Preliminary analysis of spectrum of the source in the energy range $3 - 79$ $keV$ with a thermal disk blackbody component, power law and a reflection component yielded a $kT_{in}$ of 0.628 $\pm$ 0.004 $keV$ and a $\Gamma$ of 2.40 $\pm$ 0.04 (90\% confidence errors) \citep{Tomsick2019}. MAXI J0637$-$430 was also observed in radio band with the \textit{ATCA} with flux densities of $66 \pm 15$ $\mu$Jy at 5.5 GHz and $60 \pm 10$ $\mu$Jy at 9 GHz \citep{Russell2019}. However, as the nature of the radio jet emission could not be deciphered, the source could not be properly classified using the radio/X-ray correlation in X-ray binaries. The source was also observed in infrared band with the simultaneous imaging camera \textit{SIRIUS} attached to 1.4 m telescope \textit{InfraRed Survey Facility (IRSF)}, where the estimated magnitudes in the J, H and K bands were $17.40 \pm 0.01$, $17.69 \pm 0.02$ and $17.96 \pm 0.05$, respectively \citep{Murata2019}. Spectral analysis of the X-ray data from \textit{Swift} observations with an absorbed disk blackbody model showed the source to have a $kT_{in}$ of 0.675 $\pm$ 0.003 $keV$ \citep{Knigge2019}, which is consistent with the value obtained from \textit{NuStar} observations. Since its discovery, MAXI J0637$-$430 was observed by the \textit{NICER} mission continuously with a cadence of 1-2 days, which observed it undergo a spectral state transition. After $\sim$ 23 days since its discovery, it was reported that source transitioned into the hard state \citep{Remillard2020}. Subsequent observations by \textit{Swift} showed that MAXI J0637$-$430 could possibly be approaching its quiescence level \citep{Tomsick2020}. A multi-wavelength study of the source was carried out by \cite{Tetarenko2021} using data from \textit{Swift-XRT and UVOT}, \textit{Gemini/GMOS}, \textit{ATCA} and \textit{AAVSO}. This study made use of an  irradiated accretion disk model - $\mathtt{(diskir)}$ \citep{Gierlinski2009} to derive the time-series evolution of its spectral parameters over the entire outburst cycle. Analysis of \textit{NICER} data by \citep{Jana2021} revealed the source to comprise of an ultra-soft thermal component ($kT_{in} \lesssim$ 0.6 $keV$) and a power law tail. The study also showed that its spectra do not need a thermal component corresponding to the emission from neutron star surface, thus suggesting that the compact object in the MAXI J0637$-$430 is most likely a black hole. Mass of the black hole inferred from this study is 5 $-$ 12~\(M_\odot\) for a source distance of $d < 10~kpc$ and the distance to the source is found to have a lower limit of 6.5~$kpc$. \citet{Baby2021} found that the 0.5 $-$ 25 $keV$ spectra of the source could be modelled with a multi-colour disc emission $\mathtt{(diskbb)}$ convolved with a thermal Comptonisation component $\mathtt{(thcomp)}$. Spectral fitting with the $\mathtt{kerbb}$ model in conjunction with the soft-hard transition luminosity, favour a black hole with mass between 3 $-$ 19~\(M_\odot\) and retrograde spin at a distance $<$ 15~$kpc$. Broadband spectral study on \textit{NuSTAR} data of the source showed that a two-component model, comprising of a combination of multi-color disk blackbody and thermal Comptonization component is adequate to fit the spectra only upto 10 $keV$. When higher energies are considered, scenarios involving a plunging region and reprocessing of returning disk radiation are equally possible \citep{Lazar2021}.
\\[6pt]
\noindent
Encouraged by the observation campaigns carried out by various satellite and ground based telescopes, ToO observations of MAXI J06347$-$430 were performed using the \textit{Soft X-ray Telescope (SXT)} and \textit{Large Area X-ray Proportional Counter (LAXPC)} instruments on-board \textit{AstroSat} in the 0.3 $-$ 80 $keV$ energy range on 8$^{th}$ and 21$^{st}$ November, 2019. Here, we report the results of spectral and temporal studies carried out on MAXI J0637$-$430 data from the \textit{SXT} and \textit{LAXPC} instruments. The details of \textit{AstroSat} observations and the data reduction procedures are described in Section \ref{sec:observation_datareduction}. In Section \ref{subsec:lightcurve_HID}, lightcurve and hardness intensity diagram (HID) are presented. In Section \ref{subsec:spectral_analysis}, we present the results of spectral analysis. The findings and summary of the results are discussed in Section \ref{sec:results_discussion}. 

\section{Observations and Data reduction} 
\label{sec:observation_datareduction}
\begin{figure}[ht]
    \centering
    \includegraphics[width=\columnwidth]{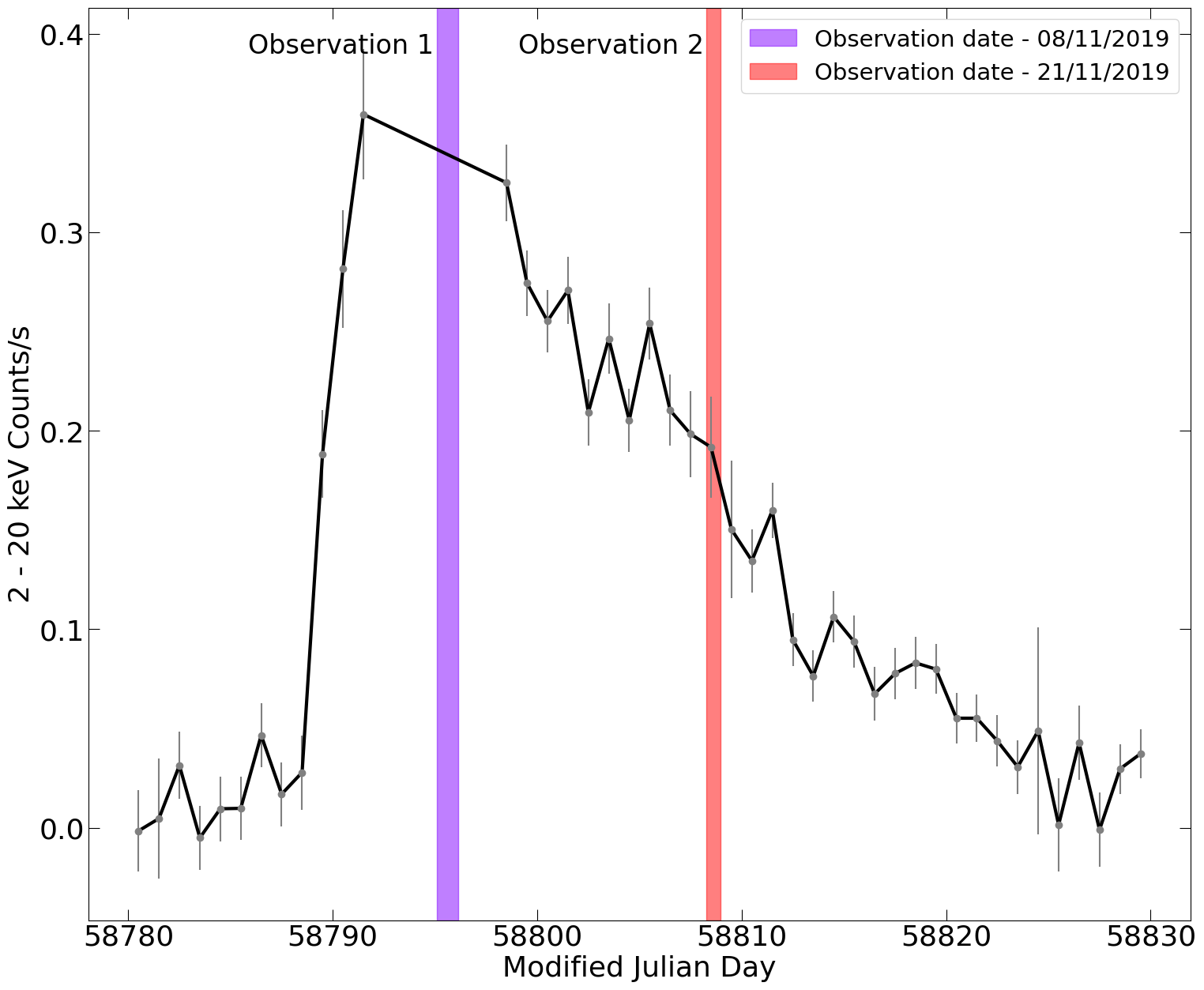}
    \caption{1 day binned MAXI lightcurve of MAXI J0637$-$430 in 2 $-$ 20 $keV$ band}
    \label{fig:MAXI_lightcurve}
\end{figure}

\noindent
ToO observations of MAXI J06347$-$430 \citep{Thomas2019} in the 0.3 $-$ 80 $keV$ energy range were carried out using \textit{SXT} and \textit{LAXPC} on-board \textit{AstroSat} for a total of $\sim$ 60 ks on 8$^{th}$ (hereafter, Observation 1), 15$^{th}$ and  21$^{st}$ November, 2019 (hereafter, Observation 2). We did not include the 15$^{th}$ November data in our study as it contains 9-pointing safety observations for the \textit{UVIT} instrument on-board \textit{AstroSat}, each with different pointing and offset. Due to this, the spectra from the individual pointing could not be combined as the effective area of the instrument changes with the offset.  Moreover, since the \textit{LAXPC} pointings were also different, flux measurement using \textit{SXT+LAXPC} data could not be made. \textit{AstroSat} observations used for our study, marked on the 2 $-$ 20 $keV$ \textit{MAXI} lightcurve in Figure \ref{fig:MAXI_lightcurve} shows that Observation 1 was carried out shortly after the outburst peak, whereas Observation 2 was performed midway during the outburst decay. The Photon Counting mode (PC) was employed for observation with \textit{SXT}, whereas for \textit{LAXPC}, the observation was carried out in the Event Analysis Mode (EA). A log of observations used for this study is given in Table \ref{Tab:Table1}. 
\\[6pt]
\textit{SXT} is a focusing telescope equipped with a Charged Coupled Device (CCD) camera that performs X-ray imaging in the 0.3 $-$ 8.0 $keV$ energy range with a spectral resolution of $\sim$ 150 $eV$ at 6 $keV$ \citep{Singh2016}. \textit{SXT} data of MAXI J06347$-$430 was processed using the standard SXT pipeline - AS1SXTLevel2-1.4b\footnote{\url{https://www.tifr.res.in/\~astrosat\_sxt/sxtpipeline.html}}. This yielded Level 2 event files for individual orbits of the observation, which were then merged into one master event file using the SXT Event Merger Tool\footnote{\label{note2}\url{https://www.tifr.res.in/\~astrosat\_sxt/dataanalysis.html}}. The merged event file was then used to extract source images with the help of XSELECT V2.4k. The source was selected between the region of 8\arcmin~ and 5\arcmin~ (inner radius) for Observations 1 and 2, respectively, and 15\arcmin~(outer radius) to reduce the effect of pile-up of the CCD. The response matrix\footnote{sxt\_pc\_mat\_g0to12.rmf} and background\footnote{SkyBkg\_comb\_EL3p5\_Cl\_Rd16p0\_v01.pha} files provided by the \textit{SXT} Payload Operations Centre (POC) were used for the analysis. Off-axis Auxiliary Response File (ARF) was created with the sxt\_ARFModule\textsuperscript{\ref{note2}}. \textit{SXT} data in the range 0.5 $-$ 5.0 $keV$ for Region 1 and 3; and 0.5 $-$ 4.8 $keV$ for Region 2  were used (Figure \ref{subsec:spectral_analysis}) as the data quality above and below these energy ranges was poor.
\\[6pt]
\textit{LAXPC} is a cluster of three co-aligned proportional counters (\textit{LAXPC-10, LAXPC-20, LAXPC-30}) that operates in the 3 $-$ 80 $keV$ energy range with an absolute temporal resolution of 10 $\mu s$ \citep{Yadav2016,Agrawal2017,Antia2017}. Data from \textit{LAXPC} was processed with the LAXPCSOFT (Format A)\footnote{\url{http://astrosat-ssc.iucaa.in/?q=laxpcData}} to obtain event files, Good Time Interval (GTI) files, lightcurves, source and background energy spectra, Response Matrix Files (RMF) and power density spectra. \textit{LAXPC-20} data alone was used for our study as it was reported by the POC that \textit{LAXPC-10} underwent an abnormal change in its gain on $28^{th}$ March, 2018 and \textit{LAXPC-30} was not operational during this time. Moreover, as the energy spectrum above 20 $keV$ was background dominated, spectral studies using \textit{LAXPC} were restricted to 4 $-$ 20 $keV$ energy range.

\section{Lightcurve and Hardness Intensity Diagram (HID)} 
\label{subsec:lightcurve_HID}
\begin{figure}[h]
    \centering
    \includegraphics[width=\columnwidth]{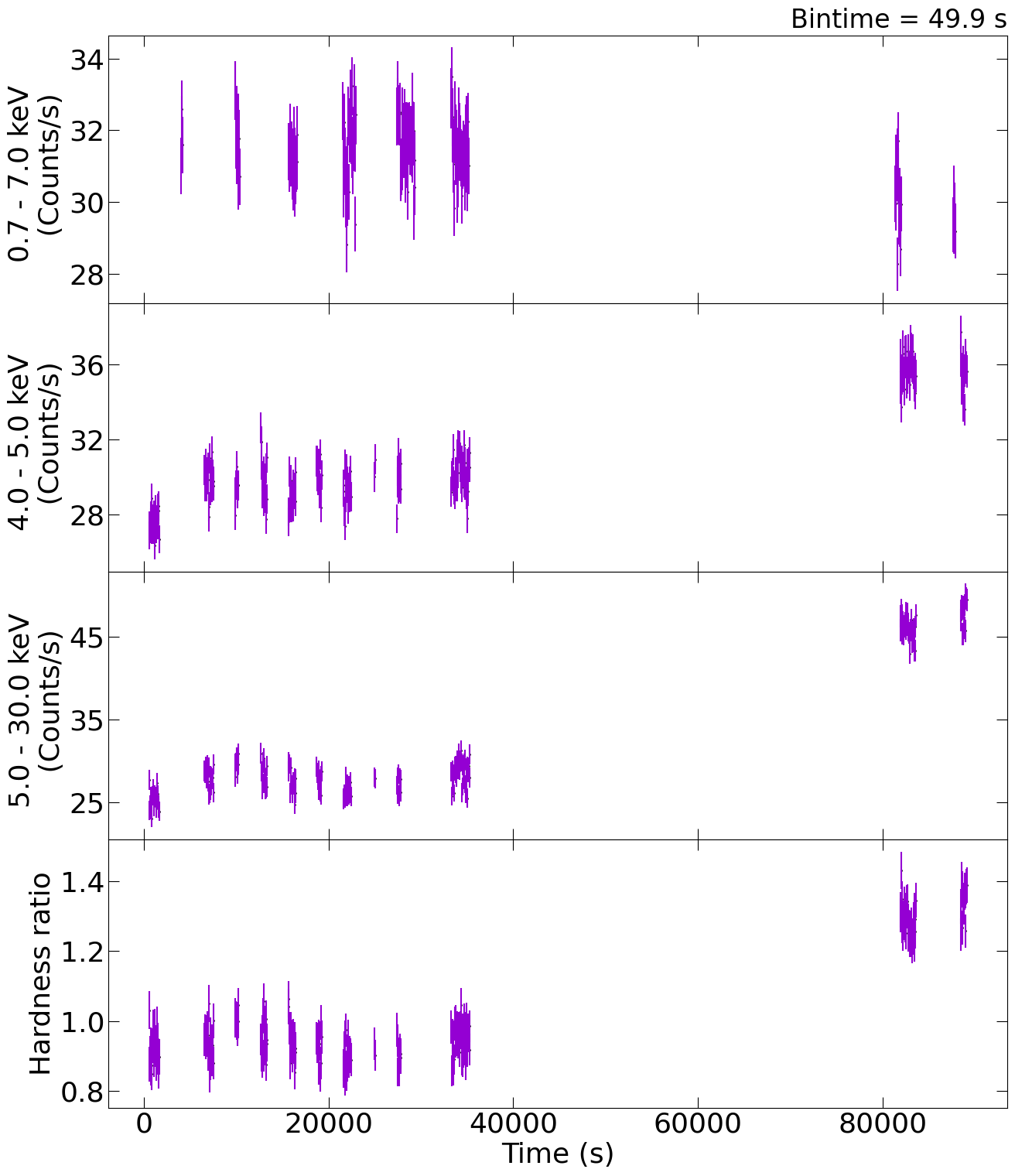}
    \caption{\textit{SXT} lightcurve of the source in 0.7 $-$ 7.0 $keV$ energy range (Panel 1),
    \textit{LAXPC-20} net lightcurve in the energy ranges 4.0 $-$ 5.0 $keV$ (Panel 2) and 5.0 $-$ 30 $keV$ (Panel 3) and their hardness ratio (Panel 4) from Observation 1}
    \label{fig:Hardness_intensity_time_147}
\end{figure}

\begin{figure}[ht]
    \centering
    \includegraphics[width=\columnwidth]{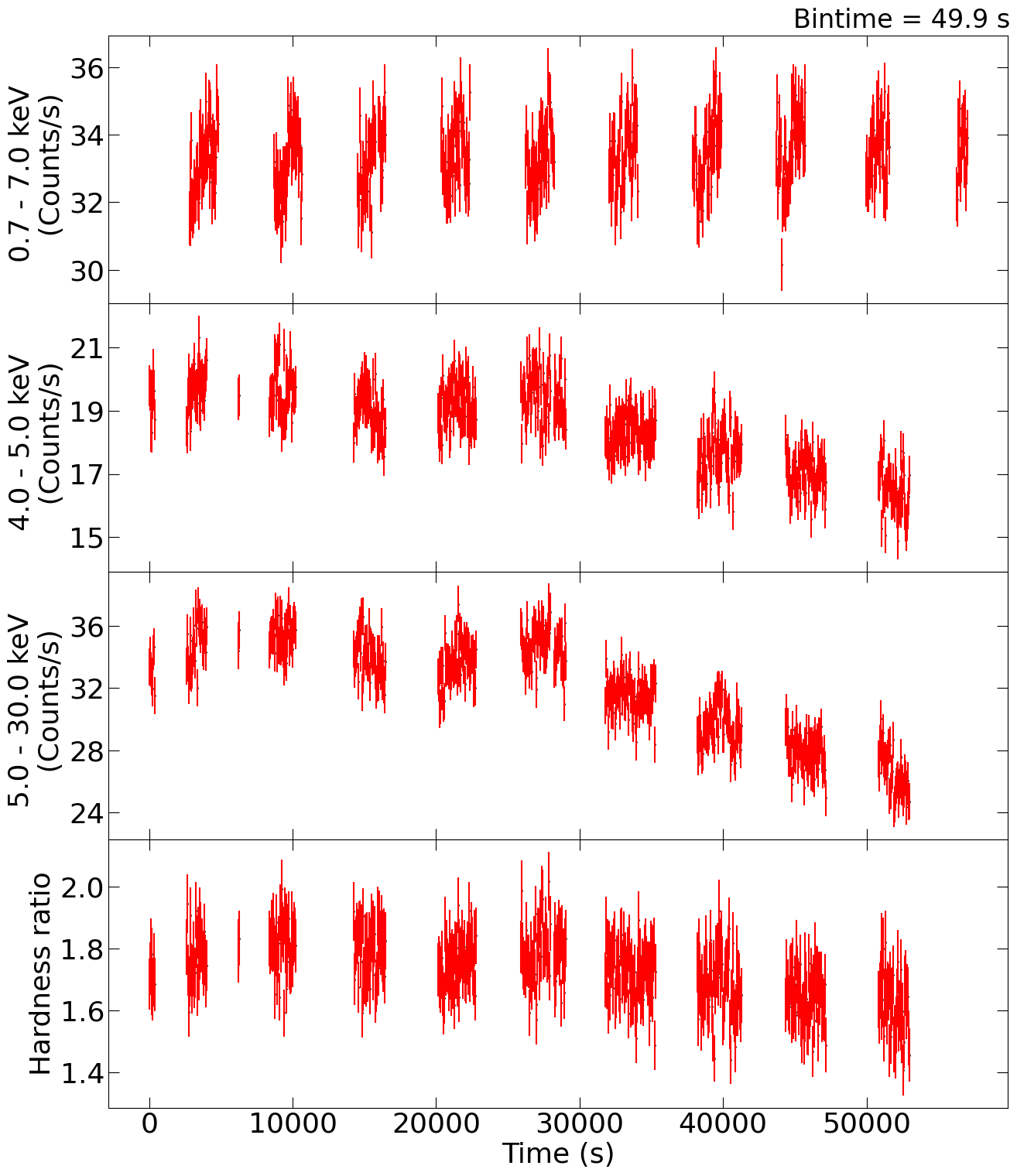}
    \caption{\textit{SXT} lightcurve of the source in 0.7 $-$ 7.0 $keV$ energy range (Panel 1),
    \textit{LAXPC-20} net lightcurve in the energy ranges 4.0 $-$ 5.0 $keV$ (Panel 2) and 5.0 $-$ 30 $keV$ (Panel 3) and their hardness ratio (Panel 4) from Observation 2}
    \label{fig:Hardness_intensity_time_159}
\end{figure}

\begin{figure}[ht]
    \centering
    \includegraphics[width=\columnwidth]{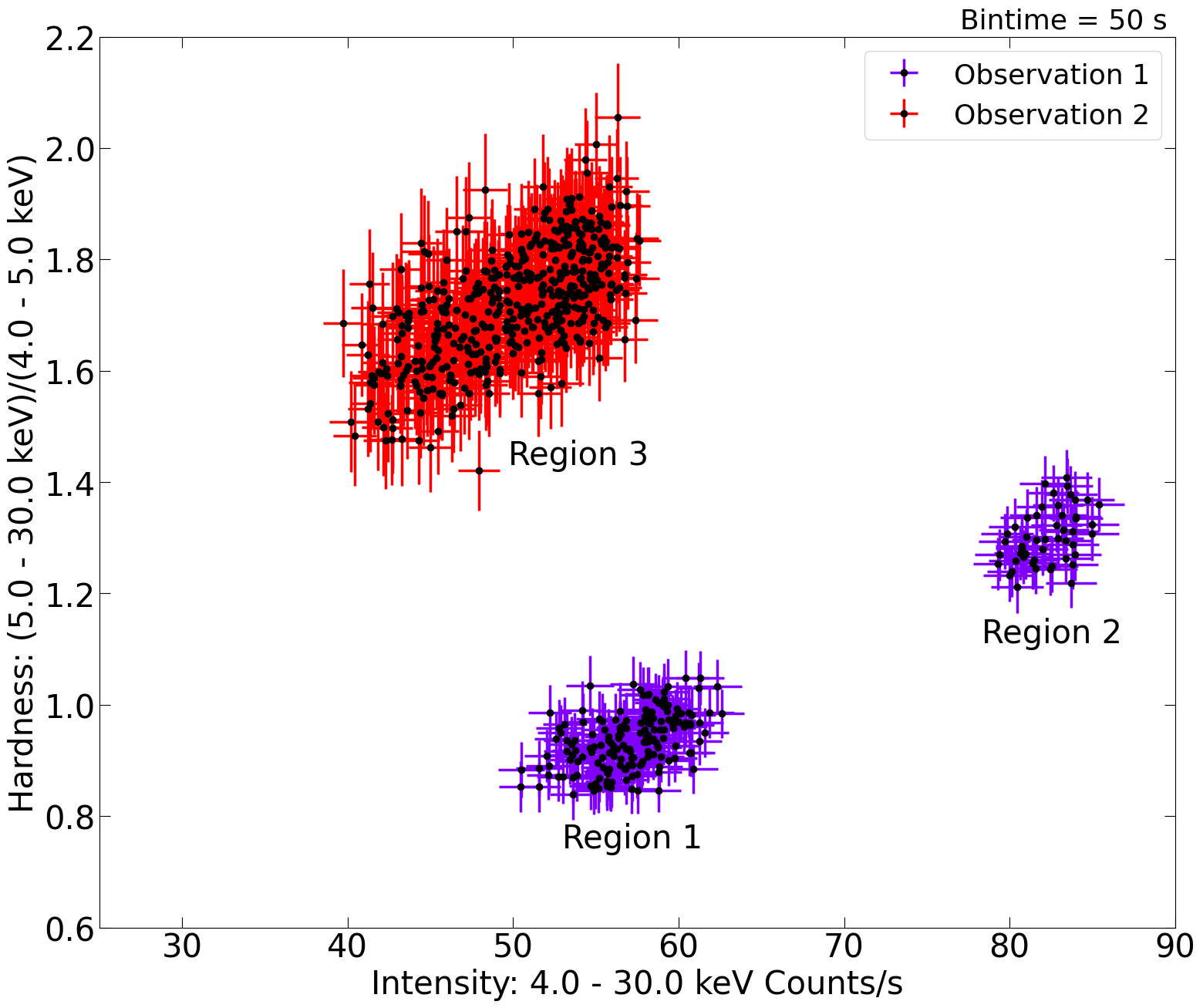}
    \caption{HID of the source from \textit{LAXPC-20} data, which is divided into 3 regions for flux resolved spectral studies.}
    \label{fig:LAXPC_HID}
\end{figure}

\begin{figure}[ht]
    \centering
    \includegraphics[width=\columnwidth]{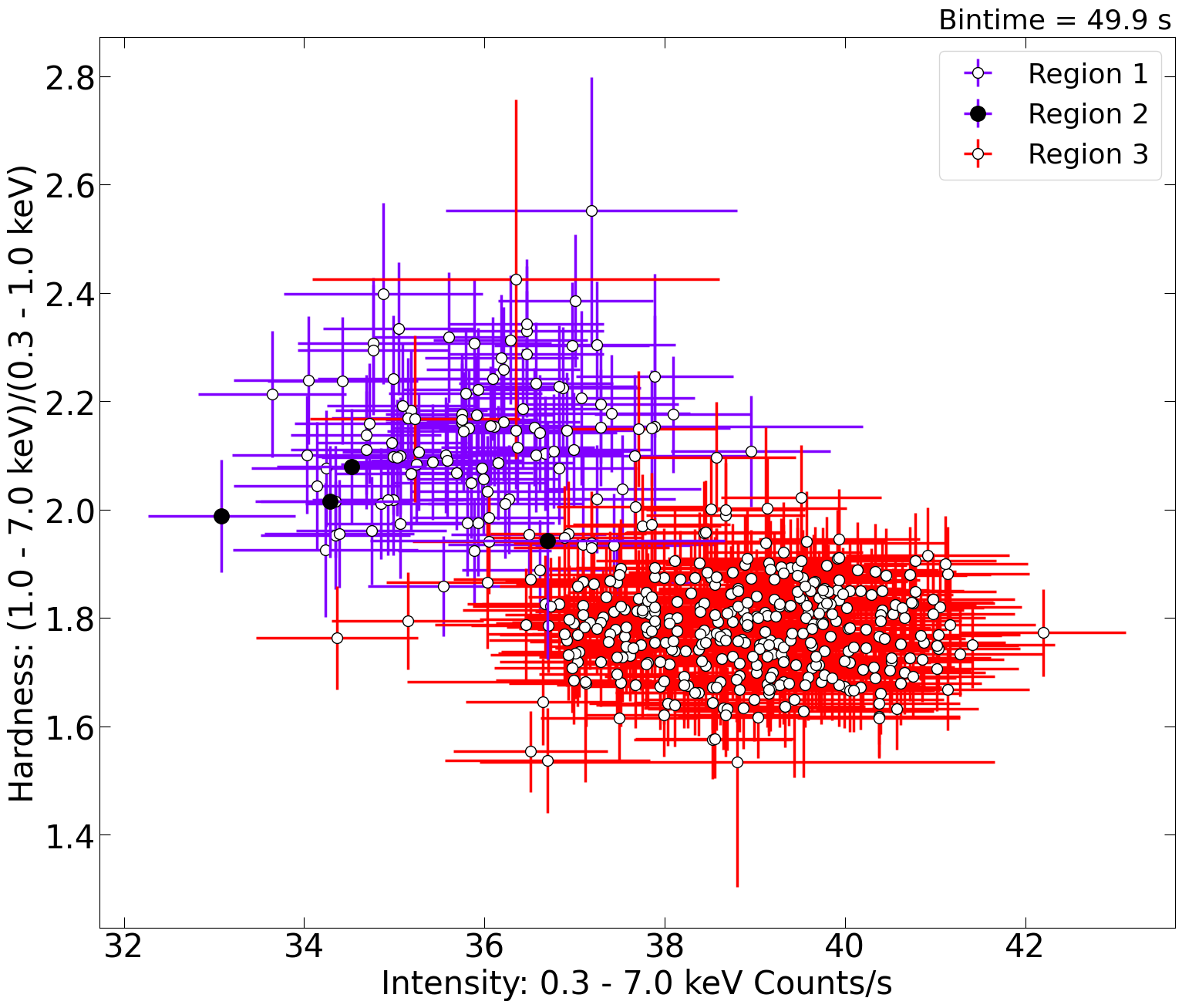}
    \caption{HID of the source from \textit{SXT} data showing Regions 1, 2 and 3 as resolved in the \textit{LAXPC}-HID.}
    \label{fig:SXT_HID}
\end{figure}

\noindent
Net lightcurves of the source were obtained in the 0.7 $-$ 7.0 $keV$ range from the SXT instrument; and 4.0 $-$ 5.0 $keV$ and 5.0 $-$ 30.0 $keV$ ranges from the \textit{LAXPC} instrument. These lightcurves were binned to $\sim$ 50 s. It is seen that during the beginning of the outburst decay i.e. in Observation 1, the source intensity remains fairly constant at $\sim$ 30 counts/s and $\sim$ 28 counts/s in the 4.0 $-$ 5.0 $keV$ (Panel 2 in Figure \ref{fig:Hardness_intensity_time_147}) and 5.0 $-$ 30 $keV$ (Panel 3 in Figure \ref{fig:Hardness_intensity_time_147}) ranges respectively. This then changes as the count rates in both energy ranges increase by a small factor towards the end of the observation. This jump in the count rate is reflected in the hardness-time diagram too (Panel 4 in Figure \ref{fig:Hardness_intensity_time_147}). In comparison, the \textit{LAXPC} net flux along with the hardness ratio is seen to decrease monotonically through the latter part of the outburst decay i.e. in Observation 2 (Panels 2, 3 and 4 in Figure \ref{fig:Hardness_intensity_time_159}). Using \textit{LAXPC-20} data of both the observations, a combined, 50 s binned HID was generated with hardness defined as the ratio of counts in the 5.0 $-$ 30.0 $keV$ range to 4.0 $-$ 5.0 $keV$ range and intensity defined as the sum of counts in the 4.0 $-$ 30 $keV$ range. From the pattern traced by the HID, it is not possible to determine if the source showed characteristics of the q-diagram exhibited by BH-LMXBs \citep{Remillard2006} or the Z or Atoll pattern exhibited by NS-LMXBs \citep{Hasinger1989}. However, it is seen that the HID showed variability in hardness from $\sim$ 0.9 to $\sim$ 1.9 (Figure \ref{fig:LAXPC_HID}). Further, in order to investigate the hardness-intensity relation in energy range $<$ 4 $keV$, data from the \textit{SXT} instrument, corresponding to the three regions in the \textit{LAXPC}-HID was used to generate an HID (Figure \ref{fig:SXT_HID}). The hardness of this \textit{SXT}-HID was defined as the ratio of counts in the 1.0 $-$ 7.0 $keV$ range to 0.3 $-$ 1.0 $keV$ range and the intensity was defined as the sum of counts in the 0.3 $-$ 7.0 $keV$ range, for time corresponding to all the three regions from the \textit{LAXPC}-HID. Regions 1 and 3 in this \textit{SXT}-HID show variations in average flux and hardness, whereas Region 2 has too less data points to make a definitive conclusion.
\begin{table*}
\centering
\caption{Observation log}
\begin{tabular}{ccccc}
\hline
Obs. ID & Date & MJD & \multicolumn{2}{c}{Exposure (ks)} \\
                           &                       &                      & SXT            & LAXPC\\
 \hline
 9000003290 & 08-11-2019            & 58795                               & 8.7           & 6.5\\
 9000003328 & 21-11-2019            & 58808                               & 19.1           & 24.8\\   
\hline
\end{tabular}
\label{Tab:Table1}
\end{table*}

\section{Spectral Analysis} 
\label{subsec:spectral_analysis}
\begin{figure*}
  \centering
  \subfloat{\includegraphics[scale=0.33]{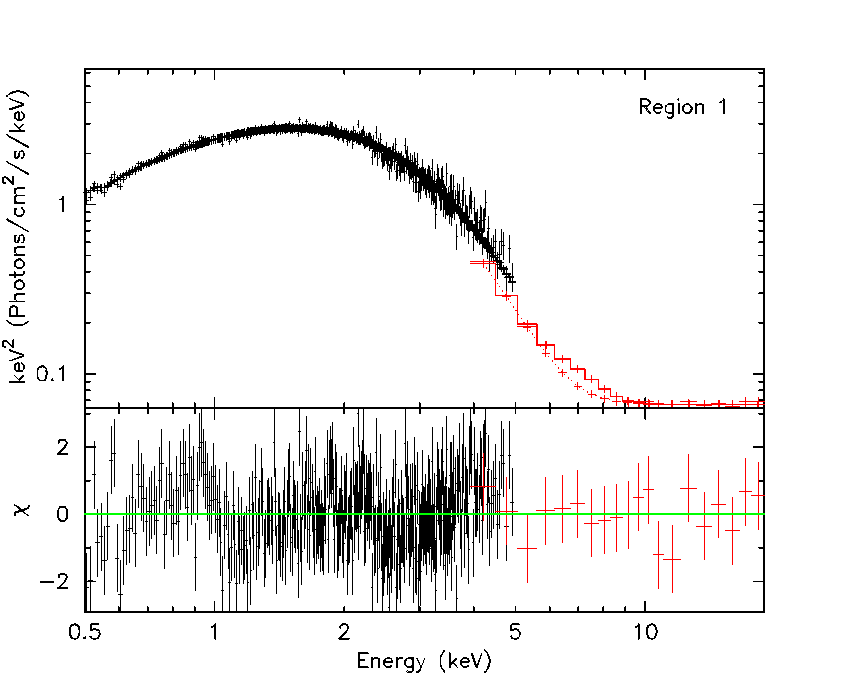}}
  \subfloat{\includegraphics[scale=0.33]{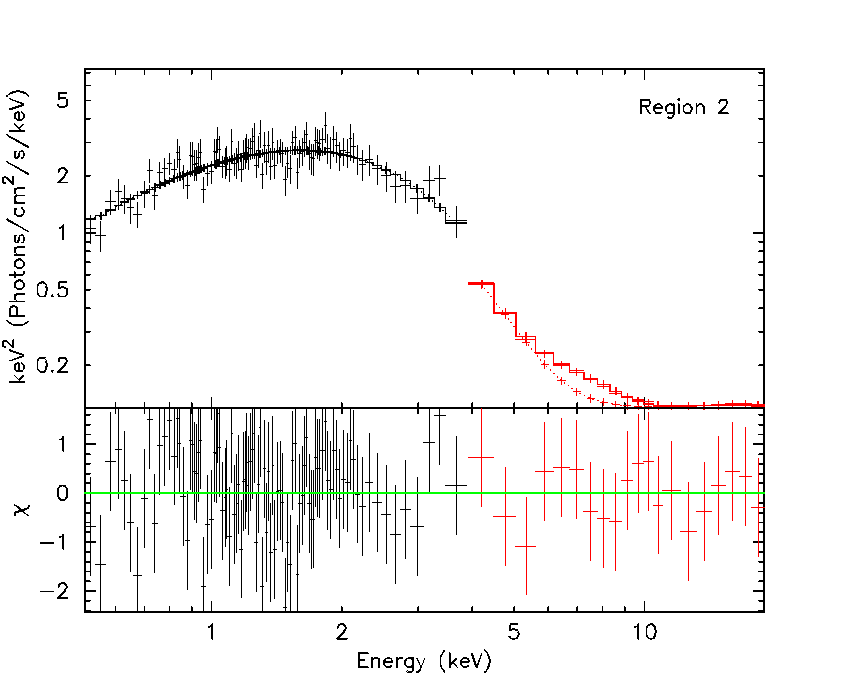}}\hspace{1em}
  \subfloat{\includegraphics[scale=0.33]{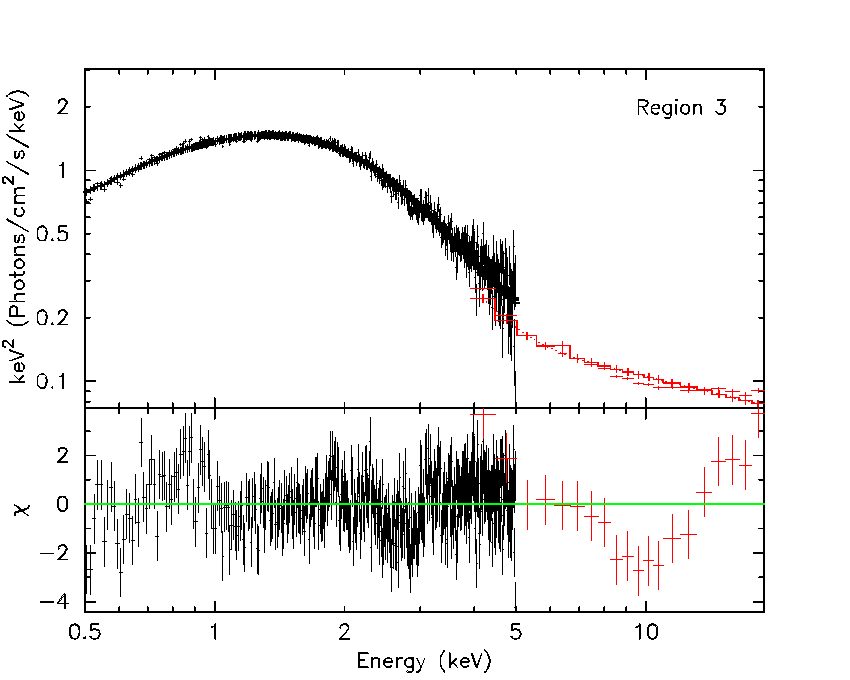}}
  \caption{SXT+LAXPC unfolded spectra of Regions 1, 2 and 3 fit with the model combination $\mathtt{constant*tbabs(simpl*diskbb+gaussian)}$. The residuals ($\chi = (data-model)/error$) are plotted in the bottom panels.}
\label{fig:Spectra}
\end{figure*}

\noindent
The simultaneous broadband X-ray spectral coverage of \textit{AstroSat} with the \textit{SXT} and \textit{LAXPC} instruments was used to perform flux resolved spectroscopy. This was done by dividing the \textit{LAXPC}-HID into 3 regions: Region 1, 2 and 3. The divisions were made such that each region denotes an isolated cluster of points in the \textit{LAXPC}-HID (Figure \ref{fig:LAXPC_HID}). Simultaneous GTIs for both the \textit{SXT} and \textit{LAXPC} instruments were generated, using which simultaneous spectra for both instruments were generated for all three regions of the \textit{LAXPC}-HID. The spectra were then fit with the spectral modelling tool $\mathtt{XSPEC}$ $\mathtt{version: 12.10.1o}$ \citep{Arnaud1996} in the energy ranges 0.5 $-$ 5.0 $keV$ from the \textit{SXT} instrument for Regions 1 and 3; and 0.5 $-$ 4.8 $keV$ for Region 2; whereas form the \textit{LAXPC} instrument, the energy range 4.0 $-$ 20 $keV$ was chosen for all three regions. These energy ranges were chosen as the spectra below 0.5 and above 4.8 $keV$ (from the \textit{SXT} instrument) showed very high residuals and that above 20.0 $keV$ (from the \textit{LAXPC} was dominated by the background. A multi-colour blackbody model - $\mathtt{diskbb}$ \citep{Mitsuda1984}, was used along with a convolution Comptonization model - $\mathtt{simpl}$ \citep{Steiner2009}, in order to explain the emission from the accretion disk and thermally Comptonized corona, respectively. The energy range for the spectral fits were extended using $\mathtt{energies}$ $\mathtt{0.01}$ $\mathtt{100}$ $\mathtt{500}$ $\mathtt{log}$ to supply an energy-binning array. To account for absorption in the interstellar medium, we used the Tuebingen-Boulder Inter-Stellar Medium absorption model - $\mathtt{tbabs}$, with the solar abundance table given by \cite{Wilms2000}. Further, a multiplicative constant factor was included to address uncertainties caused due to cross calibration of the \textit{SXT} and \textit{LAXPC} instruments. As prescribed by the POC, a systematic error of 3\% was added to all spectral fit \citep{Bhattacharya2017}. In addition to this, gain fit was performed for the \textit{SXT} data to account for the non-linear change in the detector gain. Slope of the gain was frozen to 1 leaving the offset to vary. There were positive residuals around 6.4 $keV$ indicating possible presence of a disk reflection feature. A Gaussian component with its line energy frozen at 6.4 $keV$ was later added to account for this. The addition of the Gaussian component yielded a small change in $\Delta \chi^{2}$ from 1.03 and 0.92 to 1.0 and 0.76 in Regions 1 and 2 respectively; whereas for Region 3 it remained constant. The model combination - $\mathtt{constant*tbabs (simpl*diskbb+gaussian)}$ yielded good fits for all the three regions, with $\Delta \chi^{2}$ of $\sim$ 0.9 (Figure \ref{fig:Spectra}). The best fit spectral parameters of this fit are given in Table \ref{Tab:Table2} and the corresponding spectra are give in Figure \ref{fig:Spectra}. The Norm of $\mathtt{diskbb}$ remains fairly constant in all the three regions. Hence, in order to understand the observed variation in the HID, spectral fit was repeated with the Norm of $\mathtt{diskbb}$ fixed at 2075. This value was obtained by fitting a constant through the Norms of all the three regions. Further, the unabsorbed total and disk flux were calculated in the 0.5 $-$ 20 $keV$ range using the $\mathtt{cflux}$ model. The best fit parameters of this spectral fit are given in Table \ref{Tab:Table3}. 
\\[6pt]
In addition to this, we also carried out temporal analysis with data from the \textit{LAXPC} instrument in the 4 $-$ 30 $keV$ range. However, it did not yield substantial results. 

\begin{table*}[ht]
\centering
\caption{Best fit model spectral parameters}
\begin{tabular}{lllll}
\hline
Model & Parameter & Region 1 & Region 2 & Region 3  \\
\hline
$\mathtt{tbabs}$ & $N_H (10^{20} cm^{-2})$ & 3.300 $\pm$ 0.006 & 2.120 $\substack{+0.045\\-0.021}$ & 1.250 $\pm$ 0.005\\
$\mathtt{simpl}$ & $\Gamma$ & 2.00 $\substack{+0.14\\-0.16}$ & 1.95 $\substack{+0.17\\-0.20}$ & 2.46 $\pm~0.03$ \\ 
& $FracSctr$ & 0.016 $\substack{+0.004\\-0.003}$ & 0.031 $\substack{+0.010\\-0.009}$ & 0.100 $\substack{+0.009\\-0.008}$ \\
$\mathtt{diskbb}$ & $kT_{in}$ \textit{(keV)} & 0.610~$\pm 0.005$ & 0.650 $\pm$ 0.030 & 0.520 $\pm$ 0.007 \\
 & $Norm$ & 2103 $\substack{+165\\-152}$ & 1455 $\substack{+714\\-465}$ & 2083 $\substack{+104\\-98}$ \\
$\mathtt{Gaussian}$ & $Line$ \textit{(keV)} & 6.4 (f) & 6.4 (f) & 6.4 (f) \\
 & $Width$ \textit{(keV)} & 1.06 $\substack{+0.35\\-0.40}$ & 1.49 $\substack{+0.42\\-0.47}$ & 0.20 $\substack{+0.56\\-0.37}$ \\
 & $Norm$ ($10^{-3}$) & 1.39 $\substack{+0.60\\-0.50}$ & 3.24 $\substack{+1.60\\-1.40}$ & 0.34$\pm~0.30$ \\
Reduced $\chi^2$ /dof &  & 1.03/363 & 0.76/109 & 1.33/460 \\
\hline
\end{tabular}
\label{Tab:Table2}
\end{table*}

\begin{table*}[ht]
\centering
\caption{Best fit model spectral parameters with fixed $\mathtt{diskbb}$ Norm}
\begin{tabular}{lllll}
\hline
Model & Parameter & Region 1 & Region 2 & Region 3  \\
\hline
 $\mathtt{tbabs}$ & $N_H (10^{20} cm^{-2})$ & $3.31 \pm 0.006$ & $3.87 \substack{+0.04\\-0.03}$ & $1.12 \pm 0.005$\\
 $\mathtt{simpl}$ & $\Gamma$ & $2.00 \substack{+0.14\\-0.16}$ & $2.00 \substack{+0.15\\-0.17}$ & $2.49 \pm 0.03$\\ 
&$FracSctr$ & $0.016 \substack{+0.004\\-0.003}$ & $0.028 \pm 0.007$ & $0.100 \pm 0.008$\\
 $ \mathtt{diskbb}$ & $kT_{in}~\textit{(keV)}$ & $0.610 \pm 0.003$ & $0.620 \pm 0.005$ & $0.520 \pm 0.005$\\
&$Norm$ & $2075 (f)$ & $2075 (f)$ & $2075 (f)$\\
 $\mathtt{Gaussian}$ & $Line~\textit{(keV)}$ & $6.4 (f)$ & $6.4 (f)$ & $6.4 (f)$\\
&$Width~\textit{(keV)}$ & $1.06 \substack{+0.35\\-0.40}$ & $1.32 \substack{+0.34\\-0.42}$ & $0.37 \substack{+0.56\\-0.36}$\\
&$Norm~(10^{-3})$ & $1.39 \substack{+0.60\\-0.50}$ & $3.14 \substack{+1.50\\-1.30}$ & $0.40 \substack{+0.30\\-0.20}$ \\
Unabsorbed disk flux & $(10^{-9} erg~cm^{-2}~s^{-1})$ & $5.31 \substack{+1.05\\-0.95}$ & $4.77 \substack{+1.20\\-0.84}$ & $2.58 \substack{+1.05\\-0.95}$ \\
Unabsorbed total flux & $(10^{-9} erg~cm^{-2}~s^{-1})$ & $5.53 \substack{+1.05\\-0.95}$ & $5.15 \substack{+1.18\\-0.85}$ & $3.07 \substack{+1.04\\-0.96}$ \\
Unabsorbed disk flux/total flux & & $0.96$ & $0.92$ & $0.84$ \\
$L/L_{Edd}$ & & $0.025$ & $0.023$ & $0.141$ \\ 
Reduced $\chi^2 /dof$ &  & $1.02/364$ & $0.77/110$ & $1.34/461$ \\
\hline
\end{tabular}
\label{Tab:Table3}
\end{table*}

\section{Results and Discussion} 
\label{sec:results_discussion}
In this work, we analysed the \textit{SXT} and \textit{LAXPC} data from \textit{AstroSat} observations (8$^{th}$ and 21$^{st}$ November, 2019) of MAXI J0637$-$430 in the 0.5 $-$ 20 $keV$ energy range. The analysis revealed three distinct clusters in the \textit{LAXPC}-HID of the source (Figure \ref{fig:LAXPC_HID}). However, these clusters do not form a clear pattern to give insights regarding the exact nature of the source. Monitoring the source through its entire outburst cycle with \textit{NICER} and \textit{MAXI} has revealed its HID to exhibit signatures of the various states of a BH-LMXB in outburst \citep{Jana2021,Baby2021}. Flux resolved spectral analysis showed that the spectra can be characterized by a multi-colour blackbody component arising from the accretion disk along with a thermal Comptonization component. Our choice of model - $\mathtt{diskbb}$, to characterise the soft component is in agreement with erstwhile studies carried out on the source. However, different models have been used to characterise its hard component - $\mathtt{powerlaw}$ \citep{Tetarenko2021}, $\mathtt{nthcomp}$ \citep{Jana2021,Lazar2021} and $\mathtt{thcomp}$ \citep{Baby2021}. The accretion disk temperatures, 0.61, 0.65 and 0.52 $keV$ in Regions 1, 2 and 3 respectively, point to a cool disk. This is in agreement with the studies carried out by \cite{Tetarenko2021}, \cite{Jana2021} and \cite{Baby2021}, where the disk temperature is seen to decay from $\sim$ 0.6 to $\sim$ 0.1 $keV$ during the course of the outburst. The slightly lower value of disk temperature in Region 3 correlates with the region being harder in the HID (Figure \ref{fig:LAXPC_HID}). The $\mathtt{diskbb}$ Norms of 2103, 1455 and 2083 in Regions 1, 2 and 3, indicate constant accretion disk radius throughout both the observations which is consistent with that exhibited by many BH-LMXBs in the soft state \citep{Done2007}. Similar results have been found by \citet{Baby2021} who used the convolution model $\mathtt{thcomp}$ along with $\mathtt{diskbb}$ to characterise the \textit{AstroSat} spectra of the source. Assuming a source distance of 10 $kpc$, inclination angle of 70$^{\circ}$ and a colour hardening factor of 1.7 \citep{Shimura1995}, we calculated the inner disk radius to be $\sim 98$, $\sim 81$ and $\sim 97$ $km$ in Regions 1, 2 and 3 respectively. For a black hole of $20$~\(M_\odot\), keeping the disk normalization constant at 2075, we estimated the inner disk radius to be $\sim 6$ $R_g$. This is the distance at which the Innermost Stable Circular Orbit (ISCO) is located for a non-rotating black hole. The increase in scatter fraction (obtained using $\mathtt{simpl}$ model) from 0.016 $\substack{+0.004\\-0.003}$ in Region 1 to 0.031 $\substack{+0.01\\-0.009}$ in Region 2 reflects the increased \textit{LAXPC} count rate in the HID. In addition, the presence of a Gaussian line at 6.4 $keV$ points to a reflection feature from the accretion disk. The width of this line ($\sim$ 1 $keV$ in Regions 1 and 2) shows that it is broadened due to relativistic effects around the vicinity of the central compact object. It is to be noted that the width and the Norm of the Gaussian component is significantly smaller in Region 3. The unabsorbed total flux and disk flux of the source in the 0.5 $-$ 20 $keV$ range imply Eddington fraction of 0.025, 0.023 and 0.0141 for Regions 1, 2 and 3 respectively (Table \ref{Tab:Table3}). The ratio of unabsorbed disk flux to the total flux ($\sim$ 0.9) in all three Regions (Table \ref{Tab:Table3}), suggests that the total flux is dominated by emission from the accretion disk. This consistent disk dominated flux combined with the photon power law index of $\gtrapprox$ 2, across all the three regions of the HID, shows the source to be in soft state.

\section{Conclusions} 
\label{sec:conclusion}
\noindent
We carried out flux resolved spectral studies on two observations ($\sim$ 60 ks) of the soft X-ray transient source MAXI J0637 $-$ 430 in the 0.5 $-$ 20 $keV$ energy range using the \textit{SXT} and \textit{LAXPC} instruments on-board \textit{AstroSat}. Spectral analysis of shows the source to have a cool accretion disk having temperature $\sim$ 0.55 $keV$ with a reflection feature at 6.4 $keV$. The value of photon index of $\gtrapprox 2$ and the ratio of unabsorbed disk flux to the total disk of $\sim 0.9$ points that MAXI J0637 $-$ 430 was in the soft spectral state when observed by \textit{AstroSat}. Also, it is seen that the value of the disk normalization is consistent with being constant (Table \ref{Tab:Table2}) and points to an accretion disk with an inner disk radius of 11.1 $R_g$. This is observed in several BH-LMXBs in the soft state. We conclude from our study that MAXI J0637$-$430 is a strong black hole X-ray binary candidate. Further observations and in-depth studies of the source during its future outbursts are essential to confirm its nature and unravel other physical parameters.

\section*{Acknowledgements}
\noindent
We thank the \textit{SXT} POC at TIFR, Mumbai and for the \textit{LAXPC} POC team for their support, timely release of data and providing the necessary software tools. This work has made use of software provided by HEASARC. The authors acknowledge the financial support of ISRO under \textit{AstroSat} Archival Data Utilization Program (No. DS-2B-13013(2)/9/2019-Sec.2 dated April 29, 2019). This publication uses data from the \textit{AstroSat} mission of the Indian Space Research Organisation (ISRO), archived at the Indian Space Science Data Centre (ISSDC). One of the authors (SBG) thanks the Inter-University Centre for Astronomy and Astrophysics (IUCAA), Pune for the Visiting Associateship.

\end{document}